\documentclass[lettersize,journal]{IEEEtran}
\usepackage{cite}
\usepackage{amsmath,amssymb,amsfonts}
\usepackage{graphicx}
\usepackage{textcomp}
\usepackage{algorithm}
\usepackage{algorithmic}
\usepackage{multirow} 
\usepackage{makecell}
 \usepackage{array}

 \usepackage{booktabs}
\usepackage{makecell}
\usepackage{multirow}
\usepackage{amsmath}

\usepackage{adjustbox}
\usepackage{caption}
\usepackage{subcaption}
\usepackage{color,soul}
\usepackage{siunitx}
\def\BibTeX{{\rm B\kern-.05em{\sc i\kern-.025em b}\kern-.08em
    T\kern-.1667em\lower.7ex\hbox{E}\kern-.125emX}}
\begin{document}
\title{Conditional Denoising Diffusion Model-Based Robust MR Image Reconstruction from Highly Undersampled Data}
\author{ Mohammed Alsubaie, Wenxi Liu, Linxia Gu, Ovidiu C. Andronesi, Sirani M. Perera, Xianqi Li
\thanks{Manuscript received 30 March 2025. This work was supported in part by the U.S. National Science Foundations (NSF) through the Division of Mathematical Science under Grant Numbers NSF 2410678 (X. Li) \& 2410676 (S. M. Perera) and in part by the Deanship of Graduate Studies and Scientific Research at Taif University (M. Alsubaie)}
\thanks{M. Alsubaie, W. Liu, and X. Li are with the Department of Mathematics and Systems Engineering and L. Gu is with the Department of Biomedical Engineering \& Science, Florida Institute of Technology; Melbourne, FL, 32901, USA (Email: malsubaie2019@my.fit.edu; wliu2019@my.fit.edu; xli@fit.edu; lgu@fit.edu).}
\thanks{O. C. Andronesi is with the A. A. Martinos Center for Biomedical Imaging, Department of Radiology, Massachusetts General Hospital/Harvard Medical School, Boston, MA, 02114, USA (e-mail: ANDRONESI@mgh.harvard.edu).}
\thanks{S. M. Perera is with the Department of Mathematics,
Embry-Riddle Aeronautical University, Daytona Beach, FL 32114 USA (e-mail: pereras2@erau.edu).}
\thanks{Corresponding author: Xianqi Li}}

\markboth{Journal of \LaTeX\ Class Files,~Vol.~18, No.~9, September~2020}%
{How to Use the IEEEtran \LaTeX \ Templates}

\maketitle

\begin{abstract}
Magnetic Resonance Imaging (MRI) is a critical tool in modern medical diagnostics, yet its prolonged acquisition time remains a critical limitation, especially in time-sensitive clinical scenarios. While undersampling strategies can accelerate image acquisition, they often result in image artifacts and degraded quality. Recent diffusion models have shown promise for reconstructing high-fidelity images from undersampled data by learning powerful image priors; however, most existing approaches either (i) rely on unsupervised score functions without paired supervision or (ii) apply data consistency only as a post-processing step. In this work, we introduce a conditional denoising diffusion framework with iterative data-consistency correction, which differs from prior methods by embedding the measurement model directly into every reverse diffusion step and training the model on paired undersampled–ground truth data. This hybrid design bridges generative flexibility with explicit enforcement of MRI physics. Experiments on the fastMRI dataset demonstrate that our framework consistently outperforms recent state-of-the-art deep learning and diffusion-based methods in SSIM, PSNR, and LPIPS, with LPIPS capturing perceptual improvements more faithfully. Specifically, under an acceleration factor of 8 and Gaussian 1D sampling, the proposed model achieves SSIM = $0.834\pm 0.063$, PSNR = $32.52\pm 2.63$ dB, and LPIPS = $0.063\pm 0.029$. These results demonstrate that integrating conditional supervision with iterative consistency updates yields substantial improvements in both pixel-level fidelity and perceptual realism, establishing a principled and practical advance toward robust, accelerated MRI reconstruction.
\end{abstract}


\section{Introduction}
\label{sec:introduction}
\IEEEPARstart{M}{agnetic} Resonance Imaging (MRI) plays a vital role in modern medical diagnostics \cite{b1,b2} due to its ability to offer exceptional soft tissue contrast and noninvasive visualization capabilities that are critical for detecting and monitoring a wide range of conditions \cite{b3, b4}.  However, the extended acquisition time needed to produce robust and high-quality MRI images remains a significant bottleneck, especially in clinical scenarios where rapid decision-making is crucial \cite{b5}. 

A common strategy to reduce acquisition time is k-space undersampling (or subsampling) techniques \cite{b6}, where fewer measurements are acquired during the scan. While this approach can shorten the scan time,  it introduces ill-posedness into the reconstruction process, which often results in aliasing artifacts and noise in the reconstructed images. Over the years, significant progress has been made in developing reconstruction methods for MRI from undersampled measurements. Traditional MRI reconstruction approaches, such as compressed sensing (CS) \cite{b7,b8} and total variation (TV) regularization\cite{b9,b10,b11,b12}, have been widely used to mitigate these challenges by leveraging the inherent sparsity of MRI data to improve reconstruction quality. In recent years, deep learning (DL)-based methods \cite{b13, b14, b15, b16,b17}, have shown great promise for MRI reconstruction by learning complex mappings from undersampled measurements and fully-sampled images. However, even state-of-the-art (SOTA) neural network models such as UNet variants \cite{b34,isensee2021nnu,siddique2021u} may struggle to generalize across varying sampling patterns and acceleration factors and often produce overly smooth images lacking fine anatomical detail. Moreover, their deterministic nature can limit flexibility in modeling the uncertainty inherent in ill-posed inverse problems.

More recently, diffusion-type generative models such as Denoising Diffusion Probabilistic Models (DDPMs)~\cite{b18} and score-based models~\cite{b19} have shown remarkable capability for image generation and inverse problems. Their probabilistic formulation allows for iterative denoising and improved perceptual realism compared to GANs \cite{b20}or VAEs \cite{b21,b22}. These models have inspired several recent studies in MRI reconstruction, as discussed in next Section. However, most existing diffusion-based approaches either (i) rely on unsupervised score functions without paired supervision or (ii) apply data consistency only as a post-processing step, resulting in suboptimal alignment between learned priors and measurement models. 

In this work, we propose a novel conditional denoising diffusion with enforced data consistency for robust and reliable MRI reconstruction from highly undersampled data. Our method leverages the iterative nature of diffusion models to enhance undersampled MRI reconstruction, while the introduced data fidelity term during the reverse sampling process ensures consistency between the reconstructed images and the physical constraints of the original MRI data. By reconciling the denoising diffusion model and data consistency term, our approach not only generates high-quality images but also aligns the reconstructed outputs with the actual measurements, ensuring a balance between computational reconstruction and adherence to the true underlying data. In addition to its reconstruction accuracy, the proposed framework introduces negligible computational overhead and can be readily combined with accelerated samplers for faster inference.
The contributions of this work are summarized as:
\begin{itemize}
    \item We develop a conditional denoising diffusion framework that learns to generate high-resolution (HR) MRI images from highly undersampled measurements, conditioned on an initial reconstruction.
    \item By incorporating a fidelity term into the reverse diffusion process, we enforce consistency with the physical measurement model, significantly improving reconstruction robustness.
    \item Despite being trained at a fixed acceleration factor, our model generalizes well to other sampling rates, highlighting its adaptability.
    \item Extensive experiments on the fastMRI dataset show that our method outperforms existing approaches in SSIM, PSNR, and LPIPS, with LPIPS providing a more accurate assessment of perceptual quality.

\end{itemize}

The remainder of this manuscript is organized as follows. Section~II reviews related work in MRI reconstruction and diffusion-based inverse problems. Section~III presents the proposed conditional diffusion framework with data consistency. Section~IV reports experimental results, and Section~V concludes with remarks and future directions.

\section{Related Work}
\subsection{Diffusion Models for Inverse Problems and MRI Reconstruction}
Diffusion-based generative models have recently shown exceptional ability in solving inverse problems by learning powerful image priors through iterative denoising. The foundational DDPM~\cite{b18} and the score-based stochastic differential equation (SDE) formulation~\cite{b19} provided a probabilistic view of data generation and laid the groundwork for subsequent developments in image restoration and reconstruction. Building on these ideas, several studies extended diffusion models to natural-image inverse problems, including image super-resolution \cite{b23,alsubaie2025super}, inpainting and editing \cite{b24,b25,b26}, and general linear degradations \cite{kawar2022denoising}. Later variants such as Cold Diffusion \cite{bansal2023cold} and Come-Closer-Diffuse-Faster (CCDF) \cite{chung2022come} focused on accelerating sampling and replacing Gaussian corruption with task-specific degradation processes. Collectively, these methods demonstrated that diffusion priors can achieve remarkable fidelity, stability, and diversity in reconstructing high-dimensional signals from corrupted observations.

Encouraged by these advances, researchers began exploring diffusion models in medical imaging, particularly for MRI reconstruction. Jalal et al. \cite{b27} first introduced a diffusion-prior framework for compressed-sensing MRI, showing that a pretrained denoising model can serve as a strong generative prior for reconstructing undersampled measurements. Song et al. \cite{b28} further adapted score-based diffusion to medical inverse problems, and Chung et al. \cite{b29} applied a continuous-time score-matching formulation to accelerated MRI. Comprehensive surveys such as Daras et al. \cite{daras2024survey} summarize these early explorations and highlight diffusion modeling as a promising alternative to conventional DL or variational methods.

More recent diffusion-based frameworks, such as single-posterior sampling \cite{liu2025highly} and invertible diffusion for compressed sensing \cite{chen2025invertible}, continue to extend these ideas to different acquisition settings, noise models, and network architectures. These works collectively establish diffusion as a flexible, high-fidelity generative prior for MRI reconstruction, motivating continued investigation into how such models can be further aligned with the underlying MRI physics—an aspect addressed in the next subsection.

\subsection{Data Consistency Integration in Diffusion Frameworks}
While diffusion models have demonstrated strong generative capabilities for inverse problems, their early formulations often treated the measurement model as external to the denoising process. This led to reconstructions that, although perceptually plausible, were not always consistent with acquired measurements. To address this limitation, several works have sought to incorporate data consistency mechanisms, either explicitly through optimization steps or implicitly through network conditioning, into the reverse diffusion process.

In the context of MRI reconstruction, Chung and Ye \cite{chung2025diffusion} formalized a measurement-guided diffusion process in which data-fidelity gradients were applied intermittently during sampling. Measurement-Conditioned Diffusion Models \cite{xie2022measurement} integrated conditioning on undersampled inputs, and Cycle-Consistent Bridge MRI Diffusion Models \cite{song2024self} further refined this idea using bidirectional mappings between data and image spaces. Despite these advancements, most existing approaches still rely on unsupervised score estimation and apply data-consistency enforcement either as a single projection or an external correction step after each denoising update. As a result, the learned generative prior and the physical measurement model remain only loosely coupled, which can limit convergence speed and fidelity under severe undersampling.

The proposed conditional denoising fiffusion model with enforced data consistency addresses these challenges by embedding a proximal-style fidelity correction directly into every reverse diffusion iteration. Unlike prior diffusion-based reconstruction methods that perform denoising and data consistency correction as separate alternating steps, the proposed method integrates the fidelity correction directly into each reverse diffusion iteration. This unified update jointly enforces measurement consistency and noise removal within a single sampling step, maintaining physical fidelity throughout the generative process. Trained with paired undersampled–ground-truth data, this conditional formulation ensures that the diffusion prior and the MRI measurement physics are jointly optimized. As shown in Section IV, this design leads to improved quantitative accuracy and perceptual quality while maintaining strong generalization across sampling rates.
\begin{figure*}
\begin{center}
\includegraphics[width= 0.795\textwidth]{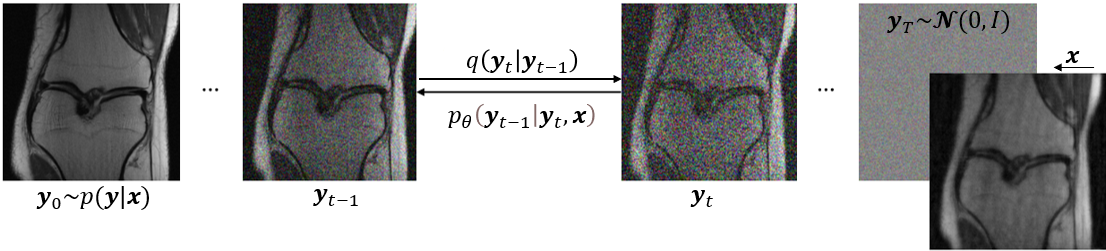}
\end{center}
\caption{The forward process (left to right) and the reverse inference process (right to left) in conditional denoising diffusion models.}
\label{fig:Fig0} 
\end{figure*} 
\section{Methods}
Building upon recent progress in diffusion-based inverse modeling and data-consistency integration, this section presents the proposed conditional denoising diffusion model with enforced data consistency for accelerated MRI reconstruction. It combines the probabilistic expressiveness of diffusion models with the physical constraints of the MRI measurement process. In contrast to existing diffusion frameworks that apply data consistency as an external post-processing step, our approach embeds the fidelity correction directly into each reverse diffusion iteration, creating a unified, physics-guided generative model.

We begin by briefly reviewing the fundamentals of conditional denoising diffusion models and the forward–reverse diffusion formulation (Section III A–D). We then describe how MRI measurement physics are incorporated into the generative process through an explicit data-consistency correction (Section III E–F). Together, these components form a hybrid reconstruction framework that leverages both learned priors and physical data constraints to achieve robust, high-fidelity image recovery from highly undersampled measurements.

\subsection{Conditional Denoising Diffusion Model}
DDPMs \cite{b23} are a class of generative models that estimate complex data distributions through a two-stage stochastic process: a forward diffusion process that gradually corrupts the data with Gaussian noise, and a reverse denoising process that learns to recover the original signal by gradually removing the noise. For undersampled MRI reconstruction, we adopt a conditional diffusion framework that learns the distribution $p(\mathbf{y}\mid\mathbf{x})$, where $\mathbf{x}$ is an undersampled image and $\mathbf{y}$ is the corresponding HR ground truth. Our objective is to transform a Gaussian noise sample into a realistic HR MRI image, guided by the contextual information present in $\mathbf{x}$.

Formally, given a dataset of paired examples $D=\{(\mathbf{x}_i,\mathbf{y}_i)\}_{i=1}^N$,  the model learns to generate samples from $p(\mathbf{y}\mid\mathbf{x})$ using a reverse diffusion process. Starting with a pure noise image at timestep $T$, i.e., $\mathbf{y}_T \sim \mathcal{N}(0, I)$ satisfying the standard normal distribution, 
the model generates a sequence of denoised images $(\mathbf{y}_{T}, \mathbf{y}_{T-1}, \dots, \mathbf{y}_0)$ using the learned distribution $p_\theta(\mathbf{y}_{t-1}\mid\mathbf{y}_{t}, \mathbf{x})$, where $t=T, T-1, \cdots, 1$. This conditional structure allows the generative process to be tightly coupled with the measured data, which improves fidelity to anatomical structures and reduces hallucinations.
To accomplish this, we train a denoising neural network model $f_{\theta}$ that takes an undersampled image  and a noisy target image as inputs and predicts the noise. The process is fully supervised and designed to recover a clean image from noisy observations across various degradation levels. A detailed description of the forward process and the associated training objective is provided in Sections II.B and II.C.



\subsection{Forward Process}
Given a HR MR image $\mathbf{y}_0$ sampled from the true data distribution $p(\mathbf{y}\mid\mathbf{x})$ (i.e. $\mathbf{y}_0 \sim p(\mathbf{y}\mid\mathbf{x})$), the forward process can be defined as a Markovian diffusion process, where Gaussian noise is incrementally added to the target image $\mathbf{y}_0$ over $T$ timesteps. 
At each timestep $t$, Gaussian noise with variance $0 < \beta_t < 1$ is added to $\mathbf{y}_{t-1}$ to produce $\mathbf{y}_t \sim q(\mathbf{y}_t \mid \mathbf{y}_{t-1})$. The non-decreasing variance $\beta_t$ can be fixed or scheduled over the $T$ timesteps. The process over the timestep is defined as 
$q(\mathbf{y}_{1:T} \mid \mathbf{y}_0 ) = \prod_{t=1}^{T} q(\mathbf{y}_t \mid \mathbf{y}_{t-1})$, 
where
$q(\mathbf{y}_t \mid \mathbf{y}_{t-1}) = \mathcal{N}(\mathbf{y}_t; \mu_t, \Sigma_t)$, 
 \( \mu_t = \sqrt{1 - \beta_t} \ \mathbf{y}_{t-1} \), and \( \Sigma_t = \beta_t I \). Defining \( \alpha_t = 1 - \beta_t \), \( \bar{\alpha}_t = \prod_{n=0}^{t} \alpha_n \), and \( \epsilon_0, \dots, \epsilon_{t-2}, \epsilon_{t-1} \sim \mathcal{N}(0, I) \), the corrupted true data at timestep $t$, i.e., $\mathbf{y}_t$ can be computed as $\mathbf{y}_t = \sqrt{\alpha_t} \mathbf{y}_{t-1} + \sqrt{1 - \alpha_t} \epsilon_{t-1}=\sqrt{\alpha_t \alpha_{t-1}} \mathbf{y}_{t-2} + \sqrt{1 - \alpha_t \alpha_{t-1}} {\epsilon_{t-2}} =\quad \dots \quad = \sqrt{\bar{\alpha}_t} \mathbf{y}_0 + \sqrt{1 - \bar{\alpha}_t} \epsilon_0$.
At any timestep $t$, the conditional distribution of $\mathbf{y}_t$ given $\mathbf{y}_0$ can be directly computed as:
\begin{equation}
q(\mathbf{y}_t \mid \mathbf{y}_0) = \mathcal{N}(\mathbf{y}_t; \mu_t = \sqrt{\bar{\alpha}_t} \mathbf{y}_0, \Sigma_t = (1 - \bar{\alpha}_t)I)
\end{equation}
As $T \to \infty$, the process converges to:
$q(\mathbf{y}_T \mid \mathbf{y}_0) \sim \mathcal{N}(0, I)$
This forward process serves two key purposes: (1) it provides noisy training targets for the denoising model, and (2) it enables stochastic training over various noise levels $\bar{\alpha}_t$ to promote stability and robustness in the learned reverse process. This process is illustrated in Fig. \ref{fig:Fig0}, from left to right. 

In addition, the posterior distribution conditioned on $\mathbf{y}_t$ and $\mathbf{y}_0$ is analytically tractable, and its Gaussian:

\begin{equation}
\label{eq:posterior}
q(\mathbf{y}_{t-1} \mid \mathbf{y}_0, \mathbf{y}_t) = \mathcal{N}(\mathbf{y}_{t-1} \mid \mu(\mathbf{y}_t, \mathbf{y}_0), \sigma^2 I),
\end{equation}
with mean and variance is given by: 
$\mu(\mathbf{y}_t, \mathbf{y}_0) = \frac{\sqrt{\alpha_t} (1 - \bar{\alpha}_{t-1})}{1 - \bar{\alpha}_t} \mathbf{y}_t + \frac{\sqrt{\bar{\alpha}_{t-1}} \beta_t}{1 - \bar{\alpha}_t} \mathbf{y}_0$, 
$\sigma^2 = \frac{(1 - \bar{\alpha}_{t-1})}{(1 - \bar{\alpha}_t)} \cdot \beta_t$.
This analytically tractable posterior distribution serves as the theoretical basis for parameterizing the reverse diffusion process. In the next section, we detail how a neural network is trained to approximate this reverse process, iteratively denoising latent variables to progressively recover the underlying data.

\subsection{Learning the Denoising Models}

The objective of the reverse process is to recover the clean image $\mathbf{y}_0$ from a noisy image $\mathbf{y}_t$ by estimating the noise added during the forward process. To achieve this, we train a neural network $f_\theta$ that predicts the noise $\epsilon$ using a noisy HR image $\tilde{\mathbf{y}} = \sqrt{\bar{\alpha}} \mathbf{y}_0 + \sqrt{1 - \bar{\alpha}} \epsilon$ with $\epsilon \sim \mathcal{N}(0, I) $  and $\bar{\alpha}$ being the sufficient statistics for the variance of the noise, and the conditioning input $\mathbf{x}$.
The objective function for training the denoising model $f_\theta$  is defined as:
\begin{equation}
\mathbb{E}_{(\mathbf{x},\mathbf{y})} \mathbb{E}_{(\epsilon,\bar{\alpha})} \| f_\theta (\mathbf{x}, \sqrt{\bar{\alpha}} \mathbf{y}_0 + \sqrt{1 - \bar{\alpha}} \epsilon, \bar{\alpha}) - \epsilon \|_p^p
\end{equation}
where $\mathbb{E}_{(\mathbf{x},\mathbf{y})}$ is the expectation over samples $(\mathbf{x},\mathbf{y})$ from the dataset, $p \in \{1,2\}$ specifies the norm, and $\bar{\alpha} \sim p(\bar{\alpha}) = \sum_{t=1}^{T} \frac{1}{T} U(\bar{\alpha}_{t-1}, \bar{\alpha}_t)$, where $U$ represents the uniform distribution. This objective encourages the model to accurately estimate the noise across a broad range of degradation levels, thereby learning a flexible denoising function applicable to diverse noise conditions. The complete training procedure is summarized in Algorithm  \ref{alg:training}.

\begin{algorithm}
\caption{Procedures for training a denoising model $f_\theta$}
\label{alg:training}
\begin{algorithmic}
    \REPEAT
        \STATE a. Sample $(\mathbf{x}, \mathbf{y}_0) \sim p(\mathbf{x}, \mathbf{y})$ 
        \STATE b. Sample $\bar{\alpha} \sim p(\bar{\alpha})$ 
        \STATE c. Sample $\epsilon \sim \mathcal{N}(0, I)$ 
        \STATE d. Compute gradient: 
        \[
        \nabla_{\theta} \mathbb{E}_{(\mathbf{x}, \mathbf{y})} \mathbb{E}_{(\epsilon,\bar{\alpha})} \| f_\theta (\mathbf{x}, \sqrt{\bar{\alpha}} \mathbf{y}_0 + \sqrt{1 - \bar{\alpha}} \epsilon, \bar{\alpha}) - \epsilon \|_p^p
        \]
    \UNTIL{convergence}
\end{algorithmic}
\end{algorithm}



\subsection{Progressive Inference through Iterative Updates}

Once the conditional denoising model $f_\theta$ is trained to predict the noise added during the forward process, inference is performed by progressively denoising the given inputs through a reverse diffusion process. More specifically, this process refines an initial Gaussian noise sample into an HR MRI image conditioned on an input $\mathbf{x}$,  as illustrated in Fig. \ref{fig:Fig0},  from right to left.
The reverse process is formulated as a Markov chain defined by:
$p_\theta (\mathbf{y}_{0:T} \mid \mathbf{x}) = p(\mathbf{y}_T) \prod_{t=1}^{T} p_\theta (\mathbf{y}_{t-1} \mid \mathbf{y}_t, \mathbf{x})$, 
$p(\mathbf{y}_T) = \mathcal{N}(\mathbf{y}_T \mid 0, I)$, and 
each transition is modeled as a Gaussian:
\begin{equation}
p_\theta (\mathbf{y}_{t-1} \mid \mathbf{y}_t, x) = \mathcal{N}(\mathbf{y}_{t-1} \mid \mu_\theta (\mathbf{x}, \mathbf{y}_t, \bar{\alpha}_t), \sigma_t^2 I)
\end{equation}
It is important to ensure that the value of $1 - \bar{\alpha}_t$ is sufficiently close to 1 so that $\mathbf{y}_T$ is approximately distributed according to the prior $p(\mathbf{y}_T) = \mathcal{N}(\mathbf{y}_T \mid 0, I)$. This guarantees that the sampling process begins with a pure Gaussian noise signal, which is desirable for the generation of target images. 


As the denoising model $f_\theta$ is trained to estimate the noise vector $\epsilon$ with any noisy HR image $\tilde{\mathbf{y}}$ (including $\mathbf{y}_t$), we can approximate $\mathbf{y}_0$ by rearranging the $\tilde{\mathbf{y}} = \sqrt{\bar{\alpha}} \mathbf{y}_0 + \sqrt{1 - \bar{\alpha}} \epsilon$ and obtain
\begin{equation}
\label{eq:y0}
\hat{\mathbf{y}}_0 = \frac{1}{\sqrt{\bar{\alpha}_t}} \left(\mathbf{y}_t - \sqrt{1 - \bar{\alpha}_t} f_\theta (\mathbf{x}, \mathbf{y}_t, \bar{\alpha}_t) \right),
\end{equation}
where we replaced $\tilde{\mathbf{y}}$ by $\mathbf{y}_t$. 
The mean of $p_\theta (\mathbf{y}_{t-1} \mid \mathbf{y}_t, \mathbf{x})$ is parameterized as:
\begin{equation}
\mu_\theta(\mathbf{x}, \mathbf{y}_t, \alpha_t) = \frac{1}{\sqrt{\alpha_t}} \left(\mathbf{y}_t - \frac{(1 - \alpha_t)}{\sqrt{1 - \bar{\alpha}_t}} f_\theta (\mathbf{x}, \mathbf{y}_t, \bar{\alpha}_t) \right)
\end{equation}
The variance of $p_\theta (\mathbf{y}_{t-1} \mid \mathbf{y}_t, x)$ is set to the default variance $\beta_t$ used in the forward diffusion process. The iterative refinement process to generate the samples takes the following form:

\begin{equation}
\mathbf{y}_{t-1} \leftarrow \frac{1}{\sqrt{\alpha_t}} \left(\mathbf{y}_t - \frac{(1 - \alpha_t)}{\sqrt{1 - \bar{\alpha}_t}} f_\theta (\mathbf{x}, \mathbf{y}_t, \bar{\alpha}_t) \right) + \sqrt{\beta_t} \epsilon_t, 
\end{equation}
where $\epsilon_t \sim \mathcal{N}(0, I)$. This formulation allows the model to reconstruct high-fidelity MRI images through a sequence of denoising operations, each informed by the underlying measurement structure and the model's learned prior. The sampling algorithm via iterative refinements is presented in Algorithm \ref{alg:Samp}.

\begin{algorithm}
\caption{Sampling via $T$ iterative refinements}
\label{alg:Samp}
\begin{algorithmic}
      \STATE Sample $\mathbf{y}_T \sim \mathcal{N}(0, I)$
      \FOR{$t = T$ to $1$}
        \STATE a. Sample $z \sim \mathcal{N}(0, I)$ if $t > 1$, else $z = 0$
        \STATE b. Compute: 
        \[
        \mathbf{y}_{t-1} \leftarrow \frac{1}{\sqrt{\alpha_t}} \left(\mathbf{y}_t - \frac{(1 - \alpha_t)}{\sqrt{1 - \bar{\alpha}_t}} f_\theta (\mathbf{x}, \mathbf{y}_t, \bar{\alpha}_t) \right) + \sqrt{\beta_t} z
        \]
      \ENDFOR
    \STATE \textbf{return} $\mathbf{y}_0$
  \end{algorithmic}
\end{algorithm}

\subsection{Physical Measurement Models for Undersampled MRI Reconstruction}

In undersampled MRI reconstruction, the physical measurement model is in the form of:
\begin{equation}
    \mathbf{b} = \mathbf{A} \mathbf{y} + \mathbf{e}
\end{equation}
where $\mathbf{b} \in \mathbb{C}^m$ is the measurement, $\mathbf{y} \in \mathbb{C}^n$ is the image to be reconstructed, $\mathbf{e} \in \mathbb{C}^m$ is the noise, and $\mathbf{A} \in \mathbb{C}^{m \times n}$ is the physical measurement operator. 
which usually takes the form $\mathbf{A} = PF S$, where $P$ represents the sampling mask, $F$ denotes the Fourier transform, and $S$ is the sensitivity map. 

In this work, we only consider the single-coil acquisition. Hence, $S$ becomes the identity matrix, leading to $\mathbf{A} = PF$. This results in the following optimization problem to solve for $\mathbf{y}$:

\begin{equation}
\label{eq:DF}
    \min_\mathbf{y} \left\{ \frac{1}{2} \| P(F(\mathbf{y})) - \mathbf{b} \|_2^2 + \lambda R(\mathbf{y}) \right\}
\end{equation}
where the first term enforces data consistency with the measured k-space data, and $R(\mathbf{y})$ is a regularization term that encodes prior knowledge about the image $\mathbf{y}$, such as sparsity or smoothness. The parameter $\lambda >0$ controls the trade-off balance between data fidelity and the regularization. 

\begin{figure*}
\begin{center}
\includegraphics[width= 0.795\textwidth]{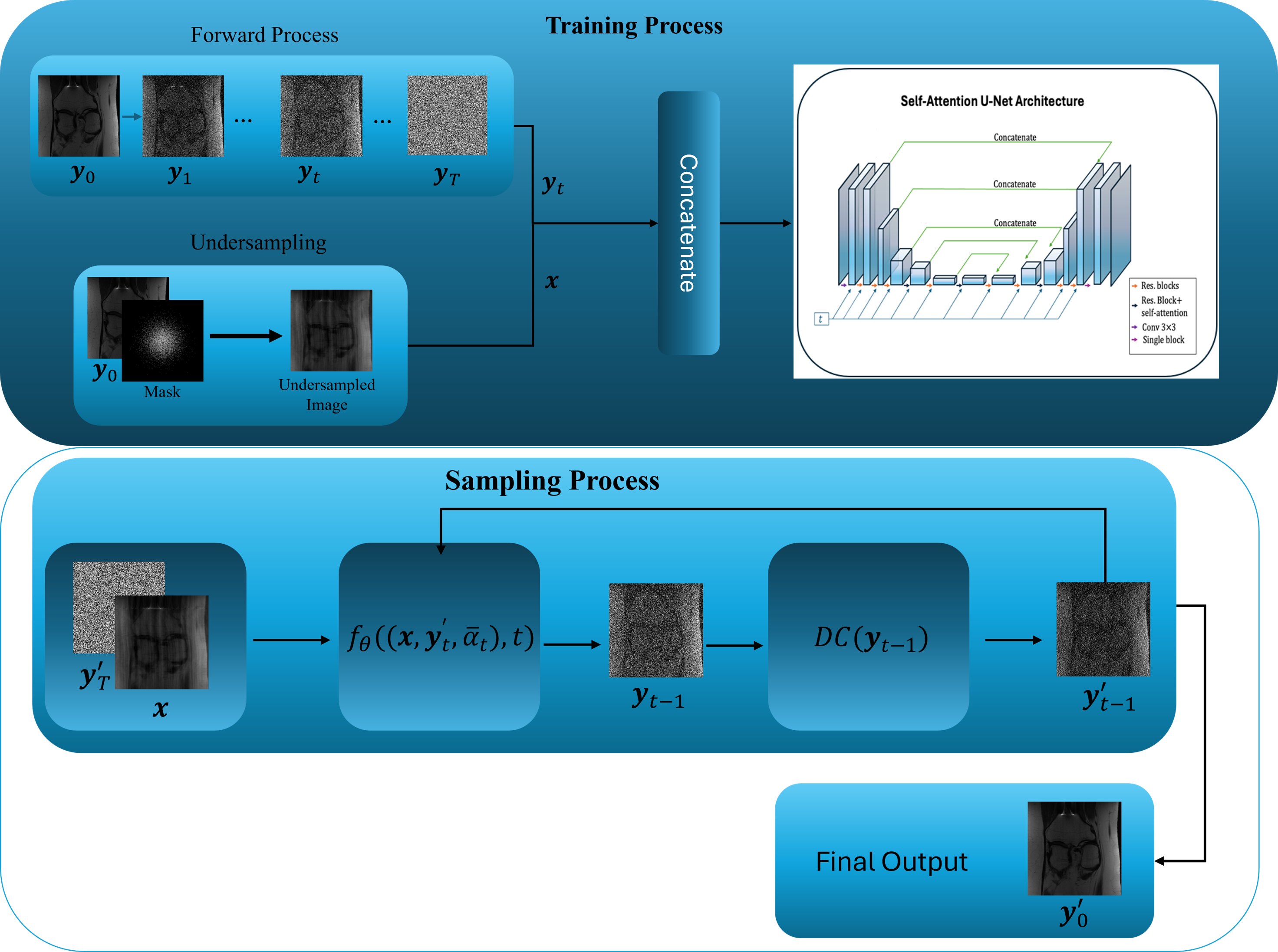}
\end{center}
\caption{Flowchart of training process and sampling process of conditional denoising diffusion model  with enforced data consistency.}
\label{fig:Fig1} 
\end{figure*} 

\subsection{Conditional Denoising Diffusion Model with Enforced Data Consistency}

In traditional approaches, $R(\mathbf{y})$ in Eq.\eqref{eq:DF}
may take the form of TV or wavelet sparsity. In our framework, however, this prior is learned implicitly via a denoising diffusion model. Rather than solving Eq. \eqref{eq:DF} through iterative optimization, we integrate data fidelity directly into the generative sampling process via a model-based correction step. This hybrid approach allows us to enforce physical consistency while maintaining the expressiveness and generalization benefits of a data-driven prior. After solving the optimization problem in Eq. \eqref{eq:DF}, we obtain:
\begin{equation}
    \mathbf{y}_{t-1} \leftarrow \frac{1}{\sqrt{\alpha_t}} \left( \mathbf{y}_t' - \frac{(1 - \alpha_t)}{\sqrt{1 - \bar{\alpha}_t}} f_\theta (\mathbf{x}, \mathbf{y}_t', \bar{\alpha}_t) \right) + \sqrt{1 - \alpha_t} \ z
\end{equation}

\begin{equation}
    \mathbf{y}_{t}' = \mathbf{y}_{t-1} - \theta \mathbf{A}^* (\mathbf{A} \mathbf{y}_{t-1} - \mathbf{b})
\end{equation}
where $\theta \in [0,1]$ and $\mathbf{A}^* $ is the conjugate transpose of $\mathbf{A}$, and hence the corresponding algorithm could be stated as Algorithm \ref{alg:SampDC}.

\begin{algorithm}
\caption{Sampling in $T$ Iterations with Data Consistency Enforced}
\label{alg:SampDC}
\begin{algorithmic}
    \STATE Initialize $\mathbf{y}_T \sim \mathcal{N}(0, I)$ and $\mathbf{y}_T' = \mathbf{y}_T$
    \FOR{$t = T$ to $1$}
        \STATE a. Sample $z \sim \mathcal{N}(0, I)$ if $t > 1$, else $z = 0$
        \STATE b. Compute:
        \[
        \mathbf{y}_{t-1} \leftarrow \frac{1}{\sqrt{\alpha_t}} \left( \mathbf{y}_t' - \frac{(1 - \alpha_t)}{\sqrt{1 - \bar{\alpha}_t}} f_\theta (\mathbf{x}, \mathbf{y}_t', \bar{\alpha}_t) \right) + \sqrt{1 - \alpha_t} z
        \]
        \STATE c. Enforce data consistency:
        \[
        \mathbf{y}_{t}' = \mathbf{y}_{t-1} - \theta \mathbf{A}^* (\mathbf{A} \mathbf{y}_{t-1} - \mathbf{b})
        \]
    \ENDFOR
    \RETURN $y_0'$
\end{algorithmic}
\end{algorithm}

\section{Experimental Settings}

\subsection{Dataset Description}

We evaluate our method using the publicly available fastMRI dataset \cite{b31}, a large-scale benchmark designed to support research in accelerated MR image reconstruction. The dataset includes raw k-space data and ground-truth images derived from multiple clinical scans. In this work, we focus exclusively on emulated single-coil (ESC) knee MR data \cite{b32}, which is synthesized from the original multi-coil acquisitions as described in \cite{b33}. This allows us to isolate the reconstruction performance of our framework without confounding effects from coil sensitivity maps. 
To train our models, we utilized both the training and validation datasets. The training dataset consisted of 973 volumes of ESC ground truth images, paired with undersampled volumes generated using various k-space masks, resulting in a total of 34,742 image pairs for training each network. Validation was performed using 199 volume pairs, corresponding to 7,135 image slices.

\subsection{Model Training and Sampling Configuration}
\begin{figure*}
\begin{center}
\includegraphics[width= 0.895\textwidth]{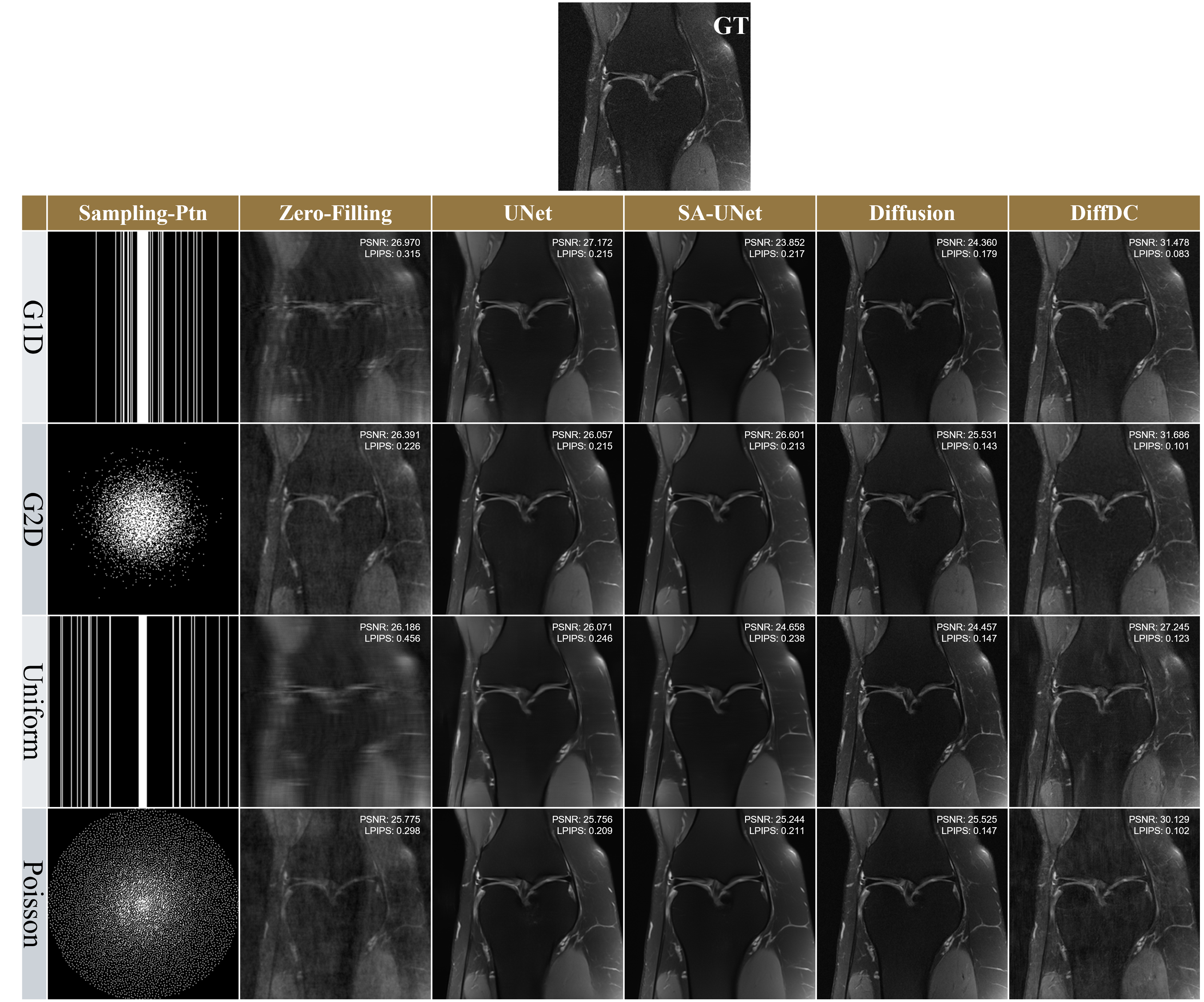}
\end{center}
\caption{MRI reconstructions with an acceleration factor of x = 8 on the same subject using models: UNet, self-attention UNet (SA-UNet), conditional denoising diffusion (Diffusion), the proposed model (DiffDC), and sampling patterns (Sampling-Ptn): Gaussian 1D (G1D), Gaussian 2D (G2D), Uniform 1D, and Poisson.}
\label{fig:Fig2} 
\end{figure*} 

All models are trained in a supervised fashion using paired undersampled and ground-truth images. Our conditional denoising diffusion model is implemented with a self-attention UNet (SA-UNet) backbone, ensuring that the total parameter count and training cost are identical. The proposed data consistency correction introduces only a lightweight projection in k-space, adding negligible computational overhead per diffusion step. Details of the SA-UNet architecture can be found in \cite{b23}. 
For noise scheduling during training, the time steps $t \sim \{0, \dots, T\}$ are uniformly sampled, and the piecewise distribution is used for sampling $\bar{\alpha}$. 
Accordingly, the model conditions directly on $\bar{\alpha}$ to allow flexibility in defining the number of diffusion steps as well as the noise schedule during inference. The maximum inference budget is set to 2000 diffusion steps. Fig. \ref{fig:Fig1} shows the flowchart of the proposed framework. 
This training pipeline was applied to four distinct DL models: Standard UNet, Self-Attention UNet (SA-UNet), conditional denoisng diffusion model (Diffusion), and diffusion model with data consistency (DiffDC). 
Statistical analysis was conducted using a test set of 30 volumes. For each volume, only slices 7 through 21 were evaluated, amounting to a total of 450 images for analysis. Because DiffDC modifies only the sampling dynamics and not the network architecture, it can be readily applied to other diffusion frameworks, including accelerated samplers such as DDIM~\cite{song2020denoising}, to achieve lower inference latency without altering reconstruction accuracy.
\begin{table*}[t] 
    \centering
    \renewcommand{\arraystretch}{1.2} 
    \begin{tabular}{c|c|ccccc}
        \hline
        \makecell{\textbf{Sampling Pattern}} & \textbf{Metric} & \textbf{Zero-Filling} & \textbf{UNet} & \textbf{SA-UNet} & \textbf{Diffusion} & \textbf{DiffDC} \\
        \hline
        \multirow{3}{*}{G1D} 
        & SSIM  & $0.693 \pm 0.059$ & $0.793 \pm 0.060$ & $0.777 \pm 0.072$ & $0.776 \pm 0.077$ & \textbf{$\mathbf{0.834 \pm 0.063}$} \\
        & PSNR  & $26.49 \pm 2.44$ & $29.22 \pm 1.59$ & $29.04 \pm 2.30$ & $28.21 \pm 2.46$ & \textbf{$\mathbf{32.52 \pm 2.63}$} \\
        & LPIPS & $0.290 \pm 0.033$ & $0.147 \pm 0.048$ & $0.140 \pm 0.052$ & $0.101 \pm 0.044$ & \textbf{$\mathbf{0.063 \pm 0.029}$} \\
        \hline
        \multirow{3}{*}{G2D} 
        & SSIM  & $0.637 \pm 0.159$ & $0.768 \pm 0.082$ & $0.785 \pm 0.083$ & $0.734 \pm 0.082$ & \textbf{$\mathbf{0.780 \pm 0.154}$} \\
        & PSNR  & $24.45 \pm 2.60$ & $28.09 \pm 1.90$ & $28.82 \pm 2.23$ & $27.45 \pm 2.20$ & \textbf{$\mathbf{29.74 \pm 2.56}$} \\
        & LPIPS & $0.246 \pm 0.041$ & $0.148 \pm 0.049$ & $0.148 \pm 0.051$ & $0.088 \pm 0.036$ & \textbf{$\mathbf{0.079 \pm 0.039}$} \\
        \hline
        \multirow{3}{*}{Uniform} 
        & SSIM  & $0.631 \pm 0.064$ & $0.744 \pm 0.069$ & \textbf{$\mathbf{0.755 \pm 0.074}$} & $0.705 \pm 0.0750$ & $0.719 \pm 0.068$ \\
        & PSNR  & $24.50 \pm 2.45$ & $27.66 \pm 1.65$ & \textbf{$\mathbf{28.21 \pm 2.00}$} & $27.02 \pm 2.07$ & $27.82 \pm 2.68$ \\
        & LPIPS & $0.420 \pm 0.039$ & $0.186 \pm 0.050$ & $0.170 \pm 0.052$ & \textbf{$\mathbf{0.098 \pm 0.035}$} & $0.108 \pm 0.029$ \\
        \hline
        \multirow{3}{*}{Poisson} 
        & SSIM  & $0.623 \pm 0.071$ & $0.770 \pm 0.077$ & $0.786 \pm 0.079$ & $0.762 \pm 0.073$ & \textbf{$\mathbf{0.799 \pm 0.057}$} \\
        & PSNR  & $23.29 \pm 2.78$ & $28.37 \pm 1.69$ & $29.22 \pm 2.24$ & $28.70 \pm 2.11$ & \textbf{$\mathbf{31.36 \pm 2.35}$} \\
        & LPIPS & $0.315 \pm 0.042$ & $0.145 \pm 0.048$ & $0.139 \pm 0.051$ & $0.086 \pm 0.037$ & \textbf{$\mathbf{0.082 \pm 0.029}$} \\
        \hline
    \end{tabular}
    \caption{Comparisons of different MRI reconstruction methods in terms of PSNR (unit in dB), SSIM, and LPIPS for an acceleration factor 8 using G1D, G2D, Uniform (1D), Poisson sampling patterns. Values are reported as mean ± standard deviation.}
    \label{tab:metrics_comparison}
\end{table*}

\begin{figure*}
\begin{center}
\includegraphics[width= 0.95\textwidth]{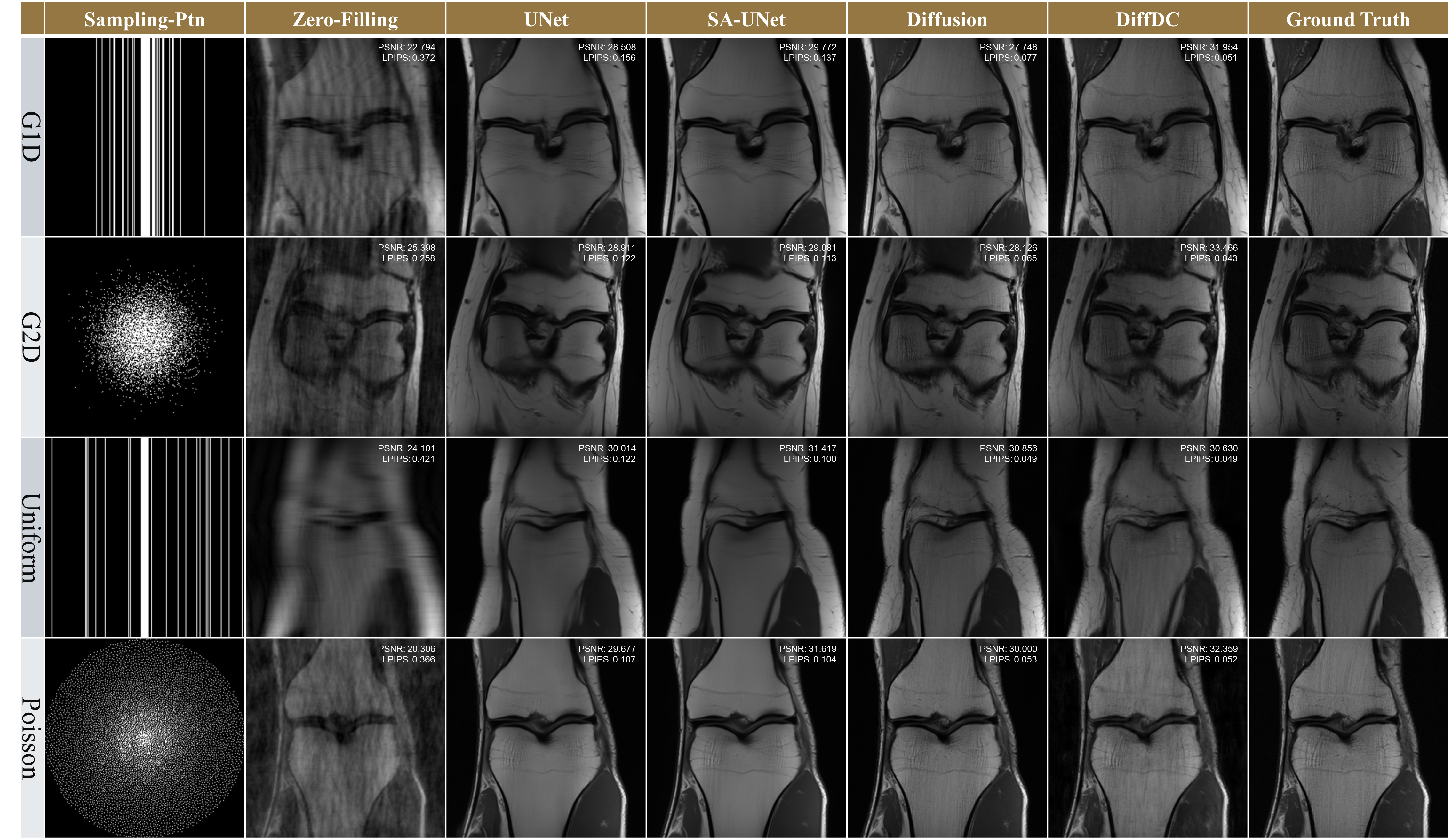}
\end{center}
\caption{MRI reconstructions with an acceleration factor of x = 8 on four different subjects using models: UNet, self-attention UNet (SA-UNet), conditional denoising diffusion (Diffusion), the proposed model (DiffDC), and sampling patterns (Sampling-Ptn): Gaussian 1D (G1D), Gaussian 2D (G2D), Uniform 1D, and Poisson. }
\label{fig:Fig3} 
\end{figure*} 
\subsection{Performance Metrics}
Our study meticulously evaluated the efficacy of advanced imaging methodologies for accelerated MRI imaging. We employed a comprehensive suite of evaluation tools, including the Structural Similarity Index (SSIM), Peak Signal-to-Noise Ratio (PSNR), and Learned Perceptual Image Patch Similarity (LPIPS) score. These metrics enabled us to assess the performance of our imaging techniques quantitatively and qualitatively. The in-depth analysis provided valuable insights into the perceptual quality and statistical accuracy of the reconstructed images, highlighting the potential of these technologies to enhance diagnostic precision in medical imaging.


\section{Numerical Results}
\subsection{Qualitative Comparison}

First, we evaluate the performance of various models on MRI reconstruction with an acceleration factor of 8, using the same undersampled fastMRI real-valued image of size $256 \times 256$. 
Fig. \ref{fig:Fig2} presents reconstructed MR images from all four models using an acceleration factor of $8\times$ under four sampling patterns: Gaussian 1D (G1D), Gaussian 2D (G2D), Uniform 1D, and Poisson. The zero-filled input images exhibit severe aliasing artifacts across all patterns.

Our proposed model, DiffDC, consistently produces reconstructions that closely resemble the ground truth, preserving fine anatomical structures with minimal artifacts. It demonstrates particularly strong performance under the G1D pattern, effectively recovering image textures that are severely lost in the input. In contrast, the Diffusion model reconstructs perceptually pleasing images but tends to blur fine details, indicating a limitation in respecting measurement fidelity. Meanwhile, UNet and SA-UNet offer acceptable denoising but frequently oversmooth structures, reducing the diagnostic quality of the output. These observations are consistent across different subjects, as shown in Fig. \ref{fig:Fig3}.

\subsection{Quantitative Evaluation}
We also provide a comprehensive comparison of quantitative metrics on the test set in Table~\ref{tab:metrics_comparison}, evaluating performance in terms of SSIM, PSNR, and average LPIPS. Across all sampling patterns, our proposed approach, DiffDC, consistently outperforms the comparison methods except the Uniform sampling one. Performance under Uniform 1D sampling shows slightly weaker gains. We attribute this to the uniform pattern’s lower incoherence, which may lead to suboptimal alignment between the diffusion prior and the projection step in data consistency enforcement. Nevertheless, DiffDC still provides competitive results under Uniform sampling. Notably, DiffDC achieves the best performance when using the Gaussian 1D sampling pattern, further highlighting its effectiveness under this configuration. These results demonstrate that the combination of diffusion-based priors and enforced data consistency enhances both pixel-level accuracy and perceptual realism.
\subsection{Ablation Study: Effect of Data Consistency Integration}
To verify that the performance gain of the proposed DiffDC model originates from data consistency mechanism rather than from network design or training differences, we performed a controlled ablation using the same SA-UNet backbone and training configuration for all diffusion variants. 

As showin in Table~\ref{tab:metrics_comparison} and Fig.~\ref{fig:Fig2} and \ref{fig:Fig3}, moving from a deterministic SA-UNet to a diffusion prior improves perceptual similarity, while adding the iterative DC correction produces a further leap in both pixel-level accuracy and visual fidelity. For example, under Gaussian 1D ($\times 8$), PNSR rises by $\approx 4 dB$ and LPIPS decreases by $\approx 0.04$ when data consistency is incorporated. These consistent gains, achieved without architectural or optimization changes, confirm that the iterative data-consistency integration is the main factor driving the superior reconstruction quality of DiffDC.

\subsection{Generalization Across Acceleration Factors}
\begin{figure}[h]
\begin{center}
\includegraphics[width= 1\linewidth]{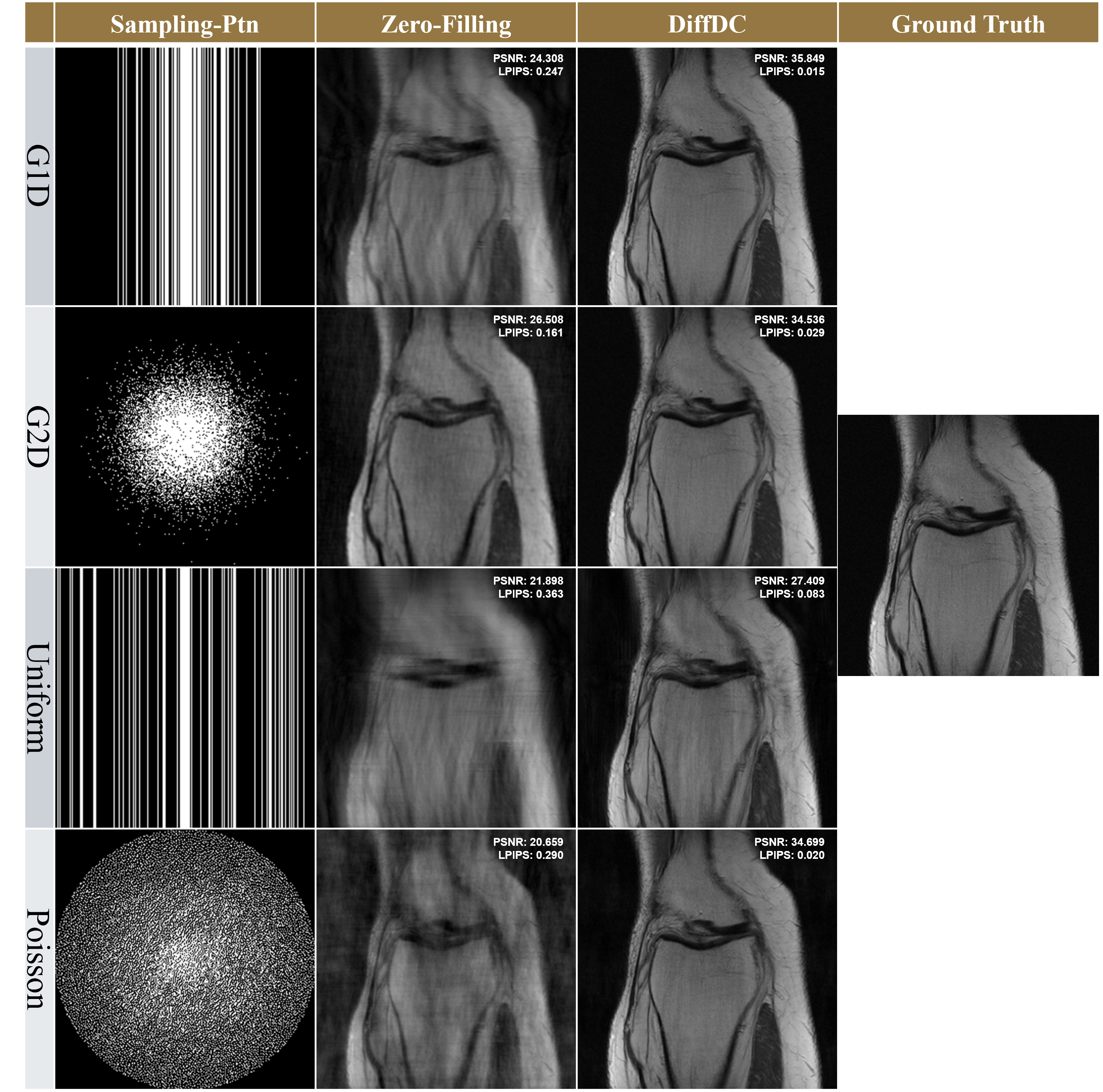}
\end{center}
\caption{MRI reconstructions with an acceleration factor of x = 4 on the same subjects by applying the proposed model (DiffDC) trained at an acceleration factor of x=8, and different sampling patterns (Sampling-Ptn): Gaussian 1D (G1D), Gaussian 2D (G2D), Uniform 1D, and Poisson. }
\label{fig:Fig4} 
\end{figure}

\begin{figure}[h]
\centering
\includegraphics[width= 1\linewidth]{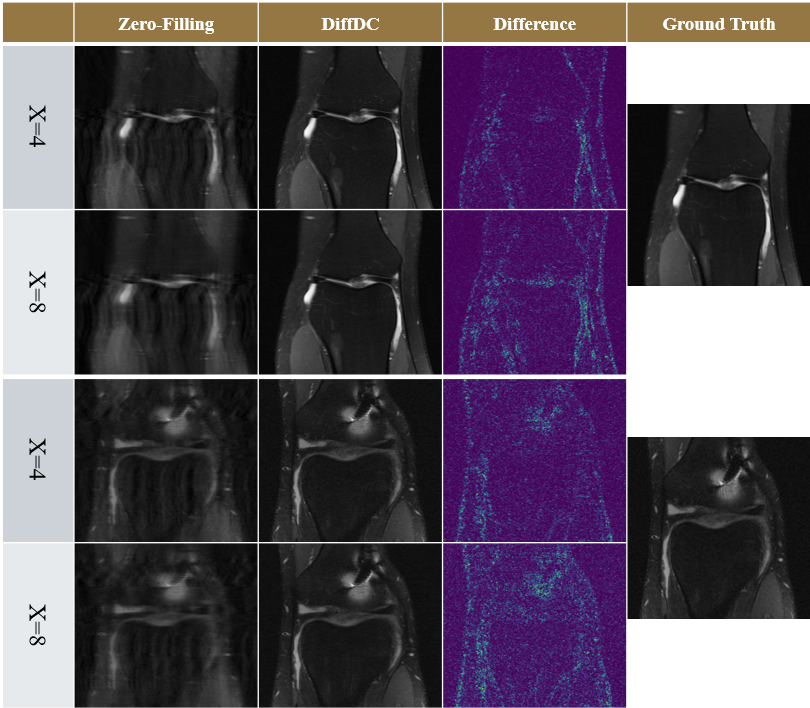}
\caption{MRI reconstructions with an acceleration factor of x = 4 using the proposed model (DiffDC) trained at an acceleration factor of x=8, and sampling patterns (Sampling-Ptn): Gaussian 1D (G1D), Gaussian 2D (G2D), Uniform 1D, and Poisson.}
\label{fig:Fig5} 
\end{figure}

To demonstrate the adaptability and generalization capability of the proposed approach, we evaluate MRI reconstruction performance under different acceleration factors. Specifically, we examine reconstruction results at an acceleration factor of 4 using various sampling patterns, even though the DiffDC model was trained exclusively at an acceleration factor of 8. As shown in Fig. \ref{fig:Fig4}, DiffDC significantly improves reconstruction quality under these conditions, highlighting its robustness to changes in undersampling rates.
This adaptability underscores the model's ability to generalize beyond its training configuration, reconstructing high-quality images with preserved details and reduced artifacts across diverse scenarios. Such versatility makes DiffDC particularly suitable for clinical applications where sampling conditions may vary dynamically. Fig. \ref{fig:Fig5} further compares the performance of the proposed approach, DiffDC, at acceleration factors of 4 and 8, showcasing the square root difference between the reconstructed images and the undersampled images. These comparisons provide additional evidence of the effectiveness and generalization capabilities of the proposed model, demonstrating its ability to deliver high-quality reconstructions across varying acceleration factors. Table \ref{tab:metrics_comparison_two_models} presents the performance metrics of the proposed DiffDC model, including PSNR, SSIM, and LPIPS. These results further highlight the effectiveness of the DiffDC model. 

\begin{table}[t] 
    \centering
    \renewcommand{\arraystretch}{1.2} 
    \begin{tabular}{c|c|cc}
        \hline
        \makecell{\textbf{Sampling Pattern}} & \textbf{Metric} & \textbf{Zero-Filling} & \textbf{DiffDC} \\
        \hline
        \multirow{3}{*}{G1D} 
        & SSIM  & $0.737 \pm 0.051$ & \textbf{$0.883 \pm 0.045$} \\
        & PSNR  & $27.83 \pm 2.35$ & \textbf{$34.65 \pm 2.38$} \\
        & LPIPS & $0.230 \pm 0.030$ & \textbf{$0.034 \pm 0.014$} \\
        \hline
        \multirow{3}{*}{G2D} 
        & SSIM  & $0.771 \pm 0.105$ & \textbf{$0.856 \pm 0.083$} \\
        & PSNR  & $30.01 \pm 3.60$ & \textbf{$33.78 \pm 4.68$} \\
        & LPIPS & $0.188 \pm 0.026$ & \textbf{$0.045 \pm 0.022$} \\
        \hline
        \multirow{3}{*}{Uniform} 
        & SSIM  & $0.653 \pm 0.063$ & \textbf{$0.815 \pm 0.056$} \\
        & PSNR  & $24.60 \pm 2.52$ & \textbf{$31.22 \pm 2.85$} \\
        & LPIPS & $0.350 \pm 0.041$ & \textbf{$0.056 \pm 0.023$} \\
        \hline
        \multirow{3}{*}{Poisson} 
        & SSIM  & $0.660 \pm 0.069$ & \textbf{$0.885 \pm 0.045$} \\
        & PSNR  & $23.68 \pm 2.71$ & \textbf{$35.47 \pm 2.47$} \\
        & LPIPS & $0.267 \pm 0.047$ & \textbf{$0.026 \pm 0.013$} \\
        \hline
    \end{tabular}
    \caption{Comparison to ground truth of the proposed method: DiffDC in terms of PSNR (unit in dB), SSIM, and LPIPS for an acceleration factor 4 using G1D, G2D, Uniform (1D), Poisson sampling patterns. }
    \label{tab:metrics_comparison_two_models}
\end{table}


\section{Conclusions}
In this work, we presented a novel conditional denoising diffusion framework with enforced data consistency for reconstructing high-quality MR images from highly undersampled data. By embedding a data fidelity term directly into the reverse diffusion process, our method establishes a principled balance between perceptual enhancement and consistency with physical measurement models. Numerical experiments on the fastMRI dataset show that the proposed DiffDC outforms SOTA methods in reconstruction performance, both quantitatively and perceptually. The model demonstrated strong adaptability and robustness, generalizing well to different acceleration factors even when trained at a fixed rate. These capabilities make DiffDC a practical and powerful solution for accelerating image reconstruction in clinical MRI without sacrificing image quality. Moving forward, we aim to explore its applicability in multi-coil acquisitions and extend its integration into real-time clinical workflows.

\section*{Acknowledgments}
This work was partially supported by the NSF grant ${\#}$2410676 \& ${\#}$2410678. An earlier version of this work appeared in the doctoral dissertation of the first author, who graduated in Dec. 2024.

\bibliographystyle{ieeetr}
\bibliography{references}

\begin{thebibliography}{10}

\bibitem{b1}
D.~B. Plewes and W.~Kucharczyk, ``Physics of mri: A primer,'' {\em Journal of Magnetic Resonance Imaging}, vol.~35, pp.~1038--1054, May 2012.

\bibitem{b2}
R.~H. Hashemi, W.~G. Bradley, and C.~J. Lisanti, {\em MRI: the basics: The Basics}.
\newblock Lippincott Williams \& Wilkins, 2012.

\bibitem{b3}
P.~M. Black, T.~Moriarty, E.~Alexander~III, P.~Stieg, E.~J. Woodard, P.~L. Gleason, C.~H. Martin, R.~Kikinis, R.~B. Schwartz, and F.~A. Jolesz, ``Development and implementation of intraoperative magnetic resonance imaging and its neurosurgical applications,'' {\em Neurosurgery}, vol.~41, no.~4, pp.~831--845, 1997.

\bibitem{b4}
M.~P. Wattjes, O.~Ciccarelli, D.~S. Reich, B.~Banwell, N.~de~Stefano, C.~Enzinger, F.~Fazekas, M.~Filippi, J.~Frederiksen, C.~Gasperini, {\em et~al.}, ``2021 magnims--cmsc--naims consensus recommendations on the use of mri in patients with multiple sclerosis,'' {\em The Lancet Neurology}, vol.~20, no.~8, pp.~653--670, 2021.

\bibitem{b5}
A.~Hosny, C.~Parmar, J.~Quackenbush, L.~H. Schwartz, and H.~J. Aerts, ``Artificial intelligence in radiology,'' {\em Nature Reviews Cancer}, vol.~18, no.~8, pp.~500--510, 2018.

\bibitem{b6}
K.~G. Hollingsworth, ``Reducing acquisition time in clinical mri by data undersampling and compressed sensing reconstruction,'' {\em Physics in Medicine and Biology}, vol.~60, pp.~R297--R322, Oct. 2015.

\bibitem{b7}
D.~L. Donoho, ``Compressed sensing,'' {\em IEEE Transactions on Information Theory}, vol.~52, pp.~1289--1306, Apr. 2006.

\bibitem{b8}
M.~Lustig, D.~Donoho, and J.~M. Pauly, ``Sparse mri: The application of compressed sensing for rapid mr imaging,'' {\em Magnetic Resonance in Medicine: An Official Journal of the International Society for Magnetic Resonance in Medicine}, vol.~58, no.~6, pp.~1182--1195, 2007.

\bibitem{b9}
L.~I. Rudin, S.~Osher, and E.~Fatemi, ``Nonlinear total variation based noise removal algorithms,'' {\em Physica D: nonlinear phenomena}, vol.~60, no.~1-4, pp.~259--268, 1992.

\bibitem{b10}
J.~Yang, W.~Yin, Y.~Zhang, and Y.~Wang, ``A fast algorithm for edge-preserving variational multichannel image restoration,'' {\em SIAM Journal on Imaging Sciences}, vol.~2, no.~2, pp.~569--592, 2009.

\bibitem{b11}
Y.~Chen, X.~Li, Y.~Ouyang, and E.~Pasiliao, ``Accelerated bregman operator splitting with backtracking,'' {\em Inverse Problems and Imaging}, vol.~11, no.~6, 2017.

\bibitem{b12}
F.~Knoll, K.~Bredies, T.~Pock, and R.~Stollberger, ``Second order total generalized variation (tgv) for mri,'' {\em Magnetic resonance in medicine}, vol.~65, no.~2, pp.~480--491, 2011.

\bibitem{b13}
I.~Oksuz, J.~Clough, A.~Bustin, G.~Cruz, C.~Prieto, R.~Botnar, D.~Rueckert, J.~A. Schnabel, and A.~P. King, ``Cardiac mr motion artefact correction from k-space using deep learning-based reconstruction,'' in {\em Machine Learning for Medical Image Reconstruction: First International Workshop, MLMIR 2018, Held in Conjunction with MICCAI 2018, Granada, Spain, September 16, 2018, Proceedings 1}, pp.~21--29, Springer, 2018.

\bibitem{b14}
R.~Shaul, I.~David, O.~Shitrit, and T.~R. Raviv, ``Subsampled brain mri reconstruction by generative adversarial neural networks,'' {\em Medical Image Analysis}, vol.~65, p.~101747, 2020.

\bibitem{b15}
S.~Yu, H.~Dong, G.~Yang, G.~Slabaugh, P.~L. Dragotti, X.~Ye, F.~Liu, S.~Arridge, J.~Keegan, D.~Firmin, {\em et~al.}, ``Deep de-aliasing for fast compressive sensing mri,'' {\em arXiv preprint arXiv:1705.07137}, 2017.

\bibitem{b16}
M.~Gaillochet, K.~C. Tezcan, and E.~Konukoglu, ``Joint reconstruction and bias field correction for undersampled mr imaging,'' in {\em International Conference on Medical Image Computing and Computer-Assisted Intervention}, pp.~44--52, Springer, 2020.

\bibitem{b17}
G.~Luo, N.~Zhao, W.~Jiang, E.~S. Hui, and P.~Cao, ``Mri reconstruction using deep bayesian estimation,'' {\em Magnetic resonance in medicine}, vol.~84, no.~4, pp.~2246--2261, 2020.

\bibitem{b34}
O.~Ronneberger, P.~Fischer, and T.~Brox, ``U-net: Convolutional networks for biomedical image segmentation,'' in {\em Medical image computing and computer-assisted intervention--MICCAI 2015: 18th international conference, Munich, Germany, October 5-9, 2015, proceedings, part III 18}, pp.~234--241, Springer, 2015.

\bibitem{isensee2021nnu}
F.~Isensee, P.~F. J{\"a}ger, P.~M. Full, P.~Vollmuth, and K.~H. Maier-Hein, ``nnu-net for brain tumor segmentation,'' in {\em Brainlesion: Glioma, Multiple Sclerosis, Stroke and Traumatic Brain Injuries: 6th International Workshop, BrainLes 2020, Held in Conjunction with MICCAI 2020, Lima, Peru, October 4, 2020, Revised Selected Papers, Part II 6}, pp.~118--132, Springer, 2021.

\bibitem{siddique2021u}
N.~Siddique, S.~Paheding, C.~P. Elkin, and V.~Devabhaktuni, ``U-net and its variants for medical image segmentation: A review of theory and applications,'' {\em IEEE access}, vol.~9, pp.~82031--82057, 2021.

\bibitem{b18}
J.~Ho, A.~Jain, and P.~Abbeel, ``Denoising diffusion probabilistic models,'' {\em Advances in neural information processing systems}, vol.~33, pp.~6840--6851, 2020.

\bibitem{b19}
Y.~Song, J.~Sohl-Dickstein, D.~P. Kingma, A.~Kumar, S.~Ermon, and B.~Poole, ``Score-based generative modeling through stochastic differential equations,'' {\em arXiv preprint arXiv:2011.13456}, 2020.

\bibitem{b20}
I.~J. Goodfellow, J.~Pouget-Abadie, M.~Mirza, B.~Xu, D.~Warde-Farley, S.~Ozair, A.~Courville, and Y.~Bengio, ``Generative adversarial nets,'' {\em Advances in neural information processing systems}, vol.~27, 2014.

\bibitem{b21}
D.~P. Kingma and M.~Welling, ``Auto-encoding variational bayes,'' in {\em International Conference on Learning Representations}, 2013.

\bibitem{b22}
A.~Vahdat and J.~Kautz, ``Nvae: A deep hierarchical variational autoencoder,'' in {\em Advances in Neural Information Processing Systems}, vol.~2020-December, 2020.
\newblock Accessed: Dec. 08, 2024.

\bibitem{b23}
C.~Saharia, J.~Ho, W.~Chan, T.~Salimans, D.~J. Fleet, and M.~Norouzi, ``Image super-resolution via iterative refinement,'' {\em IEEE transactions on pattern analysis and machine intelligence}, vol.~45, no.~4, pp.~4713--4726, 2022.

\bibitem{alsubaie2025super}
M.~Alsubaie, S.~M. Perera, L.~Gu, S.~B. Subasi, O.~C. Andronesi, and X.~Li, ``Super-resolution mr spectroscopic imaging via diffusion models for tumor metabolism mapping,'' {\em Journal of Imaging Informatics in Medicine}, pp.~1--12, 2025.

\bibitem{b24}
Y.~Song and D.~P. Kingma, ``How to train your energy-based models,'' in {\em International Conference on Learning Representations}, 2021.

\bibitem{b25}
J.~Choi, S.~Kim, Y.~Jeong, Y.~Gwon, and S.~Yoon, ``Ilvr: Conditioning method for denoising diffusion probabilistic models,'' {\em arXiv preprint arXiv:2108.02938}, 2021.

\bibitem{b26}
C.~Meng, Y.~He, Y.~Song, J.~Song, J.~Wu, J.-Y. Zhu, and S.~Ermon, ``Sdedit: Guided image synthesis and editing with stochastic differential equations,'' {\em arXiv preprint arXiv:2108.01073}, 2021.

\bibitem{kawar2022denoising}
B.~Kawar, M.~Elad, S.~Ermon, and J.~Song, ``Denoising diffusion restoration models,'' {\em Advances in neural information processing systems}, vol.~35, pp.~23593--23606, 2022.

\bibitem{bansal2023cold}
A.~Bansal, E.~Borgnia, H.-M. Chu, J.~Li, H.~Kazemi, F.~Huang, M.~Goldblum, J.~Geiping, and T.~Goldstein, ``Cold diffusion: Inverting arbitrary image transforms without noise,'' {\em Advances in Neural Information Processing Systems}, vol.~36, pp.~41259--41282, 2023.

\bibitem{chung2022come}
H.~Chung, B.~Sim, and J.~C. Ye, ``Come-closer-diffuse-faster: Accelerating conditional diffusion models for inverse problems through stochastic contraction,'' in {\em Proceedings of the IEEE/CVF conference on computer vision and pattern recognition}, pp.~12413--12422, 2022.

\bibitem{b27}
A.~Jalal, M.~Arvinte, G.~Daras, E.~Price, A.~G. Dimakis, and J.~Tamir, ``Robust compressed sensing mri with deep generative priors,'' {\em Advances in Neural Information Processing Systems}, vol.~34, pp.~14938--14954, 2021.

\bibitem{b28}
Y.~Song, L.~Shen, L.~Xing, and S.~Ermon, ``Solving inverse problems in medical imaging with score-based generative models,'' {\em arXiv preprint arXiv:2111.08005}, 2021.

\bibitem{b29}
H.~Chung and J.~C. Ye, ``Score-based diffusion models for accelerated mri,'' {\em Medical image analysis}, vol.~80, p.~102479, 2022.

\bibitem{daras2024survey}
G.~Daras, H.~Chung, C.-H. Lai, Y.~Mitsufuji, J.~C. Ye, P.~Milanfar, A.~G. Dimakis, and M.~Delbracio, ``A survey on diffusion models for inverse problems,'' {\em arXiv preprint arXiv:2410.00083}, 2024.

\bibitem{liu2025highly}
J.~Liu, Q.~Lin, Z.~Xiong, S.~Shan, C.~Liu, M.~Li, F.~Liu, G.~B. Pike, H.~Sun, and Y.~Gao, ``Highly undersampled mri reconstruction via a single posterior sampling of diffusion models,'' {\em arXiv preprint arXiv:2505.08142}, 2025.

\bibitem{chen2025invertible}
B.~Chen, Z.~Zhang, W.~Li, C.~Zhao, J.~Yu, S.~Zhao, J.~Chen, and J.~Zhang, ``Invertible diffusion models for compressed sensing,'' {\em IEEE Transactions on Pattern Analysis and Machine Intelligence}, 2025.

\bibitem{chung2025diffusion}
H.~Chung and J.~C. Ye, ``Diffusion models for inverse problems in medical imaging,'' in {\em Generative Machine Learning Models in Medical Image Computing}, pp.~129--148, Springer, 2025.

\bibitem{xie2022measurement}
Y.~Xie and Q.~Li, ``Measurement-conditioned denoising diffusion probabilistic model for under-sampled medical image reconstruction,'' in {\em International Conference on Medical Image Computing and Computer-Assisted Intervention}, pp.~655--664, Springer, 2022.

\bibitem{song2024self}
T.~Song, Y.~Wu, M.~Hu, X.~Luo, G.~Luo, G.~Wang, Y.~Guo, F.~Xu, and S.~Zhang, ``Self-consistent nested diffusion bridge for accelerated mri reconstruction,'' {\em arXiv preprint arXiv:2412.09998}, 2024.

\bibitem{b31}
F.~Knoll, J.~Zbontar, A.~Sriram, M.~J. Muckley, M.~Bruno, A.~Defazio, M.~Parente, K.~J. Geras, J.~Katsnelson, H.~Chandarana, {\em et~al.}, ``fastmri: A publicly available raw k-space and dicom dataset of knee images for accelerated mr image reconstruction using machine learning,'' {\em Radiology: Artificial Intelligence}, vol.~2, no.~1, p.~e190007, 2020.

\bibitem{b32}
J.~Zbontar, F.~Knoll, A.~Sriram, T.~Murrell, Z.~Huang, M.~J. Muckley, A.~Defazio, R.~Stern, P.~Johnson, M.~Bruno, {\em et~al.}, ``fastmri: An open dataset and benchmarks for accelerated mri,'' {\em arXiv preprint arXiv:1811.08839}, 2018.

\bibitem{b33}
M.~Tygert and J.~Zbontar, ``Simulating single-coil mri from the responses of multiple coils,'' {\em Communications in Applied Mathematics and Computational Science}, vol.~15, pp.~115--127, Nov. 2018.

\bibitem{song2020denoising}
J.~Song, C.~Meng, and S.~Ermon, ``Denoising diffusion implicit models,'' {\em arXiv preprint arXiv:2010.02502}, 2020.

\end{thebibliography}

\vspace{-1.5cm} 
\begin{IEEEbiography}{Mohammed Alsubaie} is a Ph.D student in the Department of Mathematics and Systems Engineering at Florida Institute of Technology. His research interests include optimization and deep generative AI models and their applications to medical image Analysis.
\end{IEEEbiography}

\vspace{-1.5cm}
\begin{IEEEbiography}[{\includegraphics[width=1in,height=1.25in, clip, keepaspectratio]{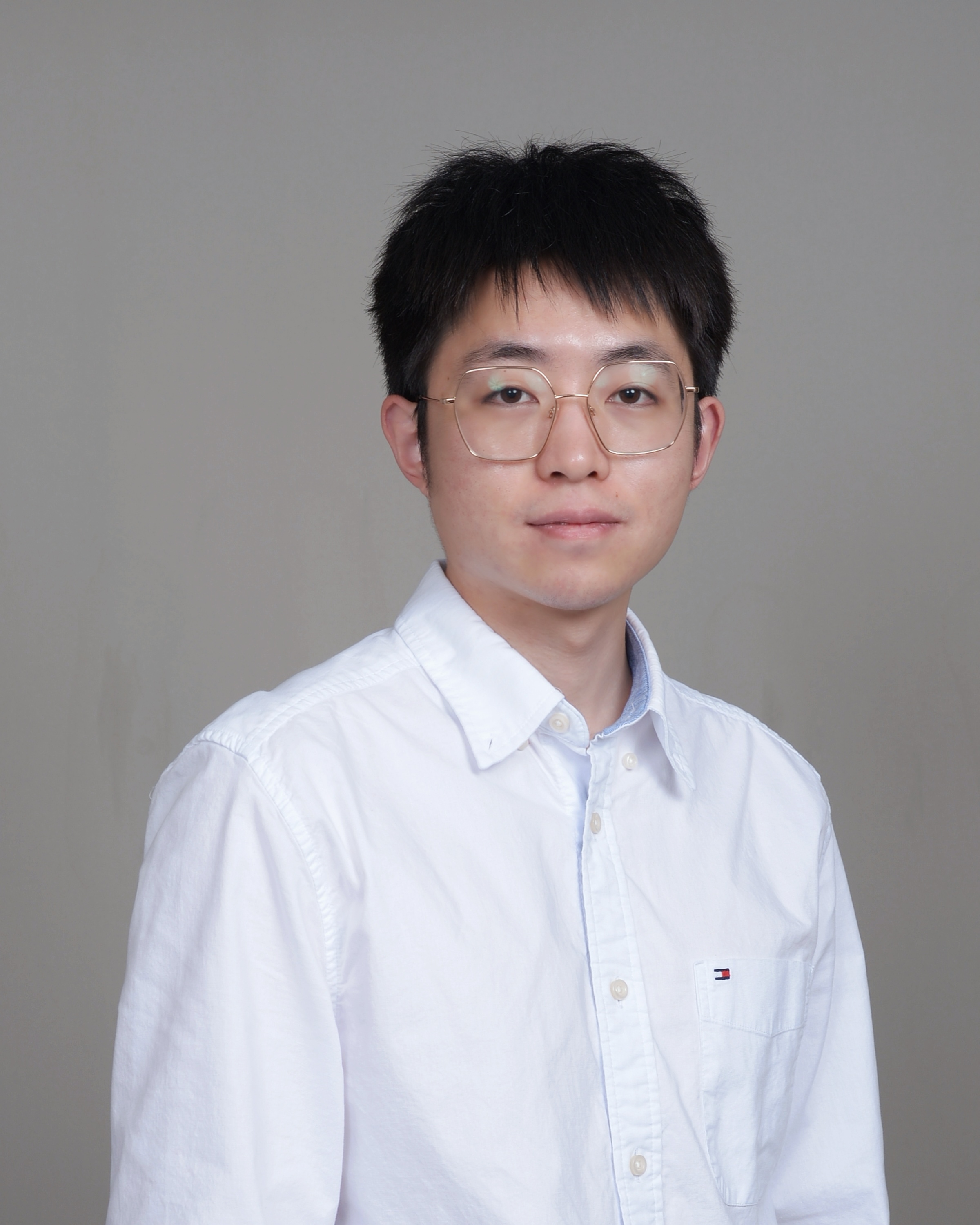}}]{Wenxi Liu} is a Ph.D student in the Department of Mathematics and Systems Engineering at Florida Institute of Technology. His research interests include digital twin modeling, stochastic optimization, and machine learning, with applications to mechanical systems and biomedical engineering.
\end{IEEEbiography}

\vspace{-1.5cm}
\begin{IEEEbiography}[{\includegraphics[width=1in,height=1.25in, clip, keepaspectratio]{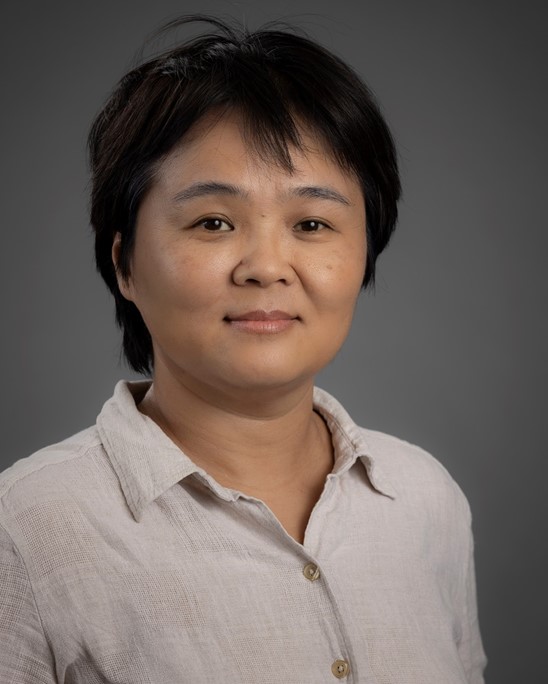}}]{Linxia Gu}(Fellow, ASME) is currently a Professor in the Department of Biomedical Engineering \&
Science, Florida Institute of Technology. Her research focuses on integrating computational mechanics, in vitro and in vivo data, and AI to predict the responses of biological systems, with applications in stented arteries, traumatic brain injury (TBI), and traumatic optic neuropathy (TON). 
Dr. Gu received the B.E. degree in Mechanical Engineering from Xi’an Jiaotong University in 1993, and the Ph.D. degree in Mechanical Engineering from the University of Florida in 2004. She has published over 150 peer-reviewed journal articles and more than 120 peer-reviewed conference papers. She is a recipient of the NSF CAREER Award and serves as an Associate Editor for Frontiers in Bioengineering and Biotechnology – Biomechanics.
\end{IEEEbiography}

\vspace{-1.5cm}
\begin{IEEEbiography}[{\includegraphics[width=1in,height=1.25in,clip,keepaspectratio]{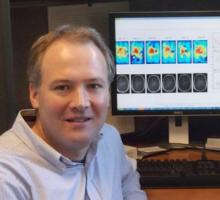}}]{Ovidiu Andronesi} is an associate Professor of Radiology at Massachusetts General Hospital, Harvard Medical School, and His research focuses on the development and clinical translation of advanced magnetic resonance imaging (MRI) and spectroscopy (MRS) techniques for neurological and oncological applications, particularly for brain tumors and neurodegenerative diseases. 
Dr. Andronesi received his Ph.D. in Physics from the Max Planck Institute for Biophysical Chemistry in Göttingen, Germany, in 2006, and his M.D. from Carol Davila University of Medicine and Pharmacy in Bucharest, Romania, in 2002. He has led numerous studies on high-field MR imaging, edited MRS, and metabolic imaging, and has authored or co-authored over 100 peer-reviewed publications.
\end{IEEEbiography}

\vspace{-1.5cm}
\begin{IEEEbiography}[{\includegraphics[width=1in,height=1.25in, clip, keepaspectratio]{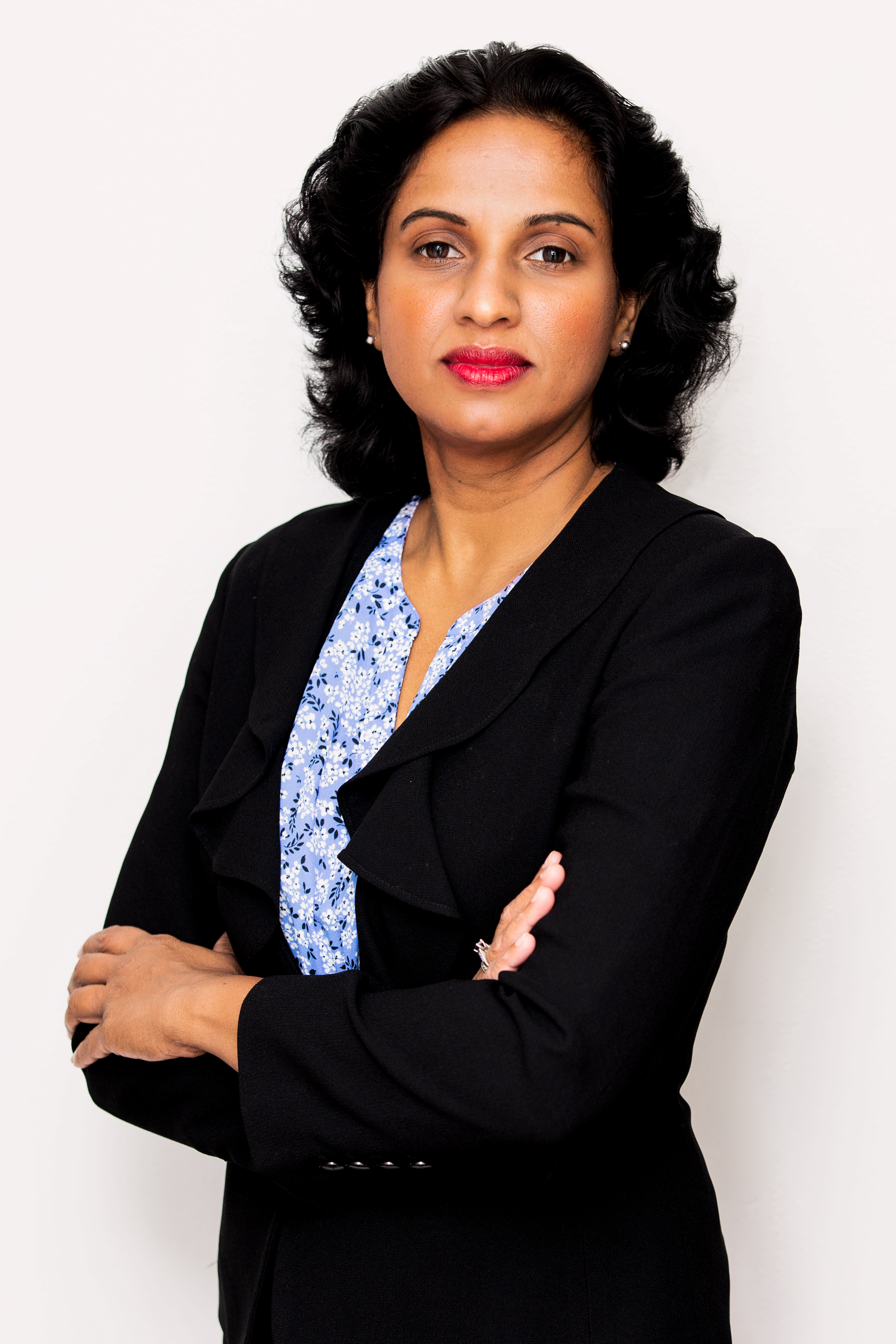}}]{Sirani M. Perera} received a B.S.(First Class Honors) in Mathematics from the University of Sri Jayewardenepura, Sri Lanka, in 2004. She became the first Sri Lankan mathematics undergraduate to be selected by the University of Cambridge’s Center for Mathematical Sciences in the United Kingdom with a full Cambridge Commonwealth Trust scholarship. She received her Master of Advanced Studies (MASt with Honors) in Mathematics from
the University of Cambridge, United Kingdom, in 2006. From 2007 to 2008, she was a lecturer in Mathematics at the University of Colombo, Sri Lanka, and she earned a Ph.D. degree in Mathematics from the University of Connecticut, USA, in 2012. 

Later, she joined Embry-Riddle Aeronautical University (ERAU) and has been working as an Associate Professor of Mathematics. Her research interests include applied linear algebra, computational mathematics, and scientific computing, along with the development of low-complexity classical and machine learning algorithms. She has authored over 40 journal articles, book chapters, and media publications, and is the inventor of the patent titled {\it Reduced Multiplicative Complexity Discrete Cosine Transform (DCT) Circuitry} granted a patent by the USPTO on January 26, 2021. 

\end{IEEEbiography}

\vspace{-1.5cm}
\begin{IEEEbiography}[{\includegraphics[width=1in,height=1.25in, clip, keepaspectratio]{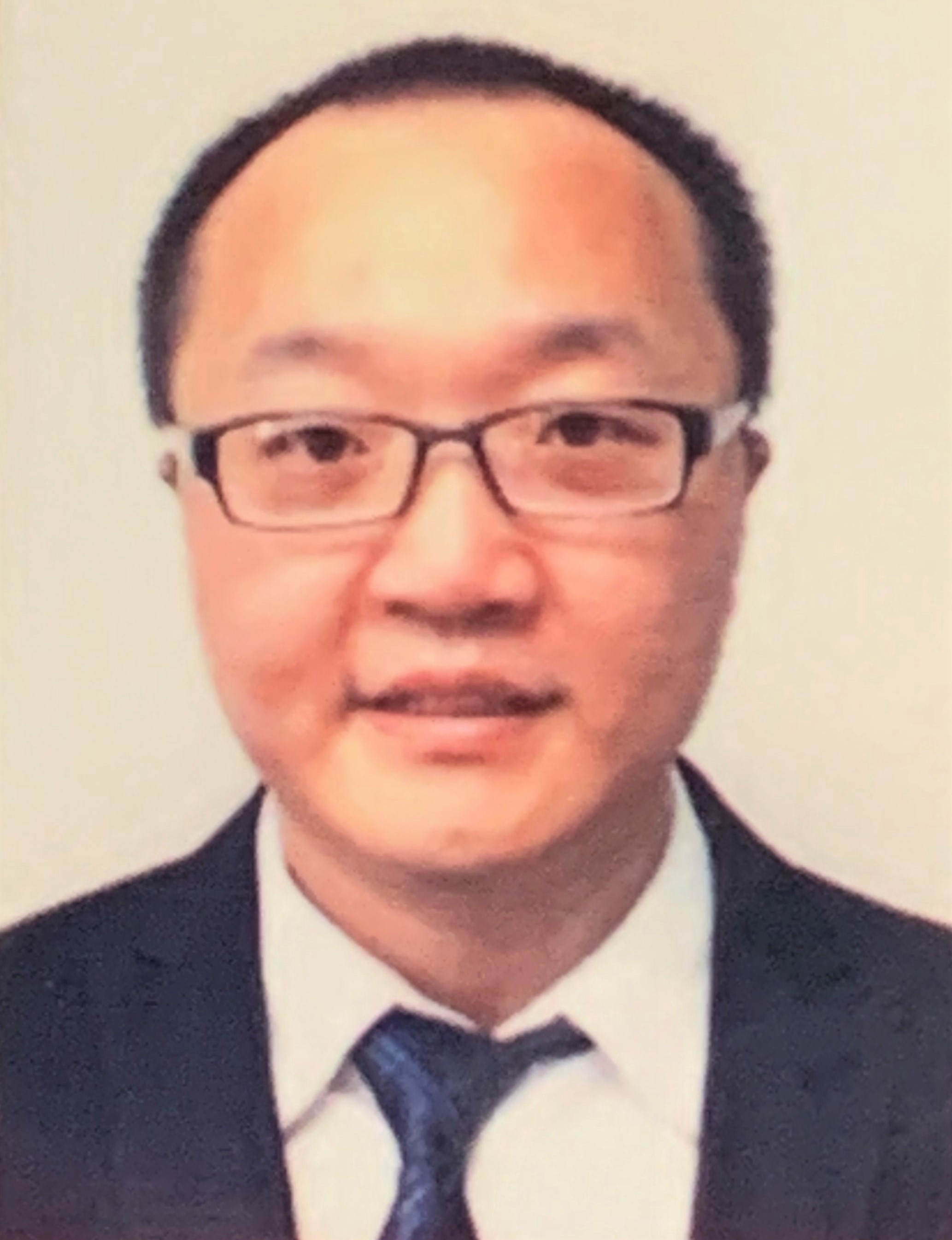}}]{Xianqi Li} is an Assistant Professor in the Department of Mathematics and Systems Engineering at Florida Institute of Technology. His research interests include optimization, machine/deep learning, and data analytics, focusing on developing models and algorithms for imaging informatics and data analytics. Previously, Dr. Li was a postdoctoral research fellow at Mass General Hospital, Harvard Medical School (2018-2021). He earned his Ph.D. in Applied Mathematics from the University of Florida in 2018, with a thesis on optimization methods for inverse problems, and an M.S. in Electrical and Computer Engineering from the same institution in 2015.
\end{IEEEbiography}
\end{document}